\documentclass[aps,twocolumn,nofootinbib,superscriptaddress,amsfonts,floatfix
]{revtex4-1} 
\usepackage{graphicx,amsmath,amssymb,amstext}
\usepackage{amssymb,amsbsy,amsfonts,amsthm,color}

\usepackage{mathtools,nccmath}

\usepackage{epsfig}
\usepackage{graphicx}
\usepackage{subfigure}
\usepackage[dvipsnames]{xcolor}
\definecolor{linkcolor}{rgb}{0.0,0.3,0.5}
\usepackage[unicode, colorlinks=true, linkcolor=linkcolor, citecolor=linkcolor, 
filecolor=linkcolor,urlcolor=linkcolor, pdfusetitle]{hyperref}
\usepackage{cancel}
\usepackage{balance}
\usepackage[capitalise]{cleveref}
\usepackage{xspace}
\usepackage{enumerate}
\usepackage{ulem}
\usepackage{mdframed}
\usepackage{booktabs}
\usepackage{makecell}
\usepackage{enumitem,tabularx,multirow}
\usepackage{aas_macros}
\usepackage{colortbl}
\usepackage{orcidlink}

\AtBeginDocument{%
  \heavyrulewidth=.08em
  \lightrulewidth=.05em
  \cmidrulewidth=.03em
  \belowrulesep=.65ex
  \belowbottomsep=0pt
  \aboverulesep=.4ex
  \abovetopsep=0pt
  \cmidrulesep=\doublerulesep
  \cmidrulekern=.5em
  \defaultaddspace=.5em
}

\graphicspath{{Figures/}}

\usepackage{color}

\newcommand{\Beq}{\begin{eqnarray}}
\newcommand{\Eeq}{\end{eqnarray}}

\def\lsim{\mathrel {\vcenter {\baselineskip 0pt \kern 0pt \hbox{$<$} \kern 0pt \hbox{$\sim$} }}}

\def\gsim{\mathrel {\vcenter {\baselineskip 0pt \kern 0pt \hbox{$>$} \kern 0pt \hbox{$\sim$} }}}

\newcommand{\RomanNumeralCaps}[1]

\definecolor{mypurple}{RGB}{143, 116, 210}

\def\-{\,-\,}
\def\={\,=\,}
\def\+{\,+\,}

\definecolor{burgundy}{rgb}{0.5, 0.0, 0.13}
\definecolor{coolblack}{rgb}{0.0, 0.18, 0.39}
\definecolor{darkblue}{rgb}{0.0, 0.0, 0.55}
\definecolor{darkgreen}{rgb}{0.0, 0.2, 0.13}

\begin{document}
\title{Medium Effects in MIT Bag Model for quark matter: Self consistent thermodynamical  treatment  }

\author{Suman Pal,\orcidlink{0009-0000-5944-4261}}
\email{sumanvecc@gmail.com}
\affiliation{Physics Group, Variable Energy Cyclotron Centre, 1/AF Bidhan Nagar, Kolkata 700064, India}
\affiliation{Homi Bhabha National Institute, Training School Complex, Anushakti Nagar, Mumbai 400085, India}

\author{Gargi Chaudhuri,\orcidlink{0000-0002-8913-0658}}
\email{gargi@vecc.gov.in}
\affiliation{Physics Group, Variable Energy Cyclotron Centre, 1/AF Bidhan Nagar, Kolkata 700064, India}
\affiliation{Homi Bhabha National Institute, Training School Complex, Anushakti Nagar, Mumbai 400085, India}

\begin{abstract}

The study of strange quark matter  within the framework of the density-dependent MIT Bag model using the Grand Canonical ensemble is thermodynamically inconsistent. In this work, it is shown  that if the medium effects are incorporated through  a density-dependent Bag pressure in the Grand Canonical ensemble, then the Euler relation is violated.  If  Euler relation is used then the minimum of energy per baryon  does not occur at zero pressure. In order to overcome this inconsistency,   we propose the medium effect of the strange quark matter in the form of  chemical potential dependent Bag pressure in the grand Canonical ensemble. The density dependent Bag pressure which has been used in Grand Canonical ensemble so far can however be used in Canonical ensemble without violating the laws of thermodynamics. These prescriptions will obey the Euler relation as well as the minimum energy per baryon will coincide with the zero of pressure and hence can be considered to be self consistent. These equations of state in the Grand Canonical ensemble can be further used to construct the Mass-Radius and other structural properties of the  strange quark stars as well as hybrid stars. In our present work we have calculated the mass radius diagram of strange stars only using this formalism.

\end{abstract}
\maketitle


\section{Introduction} \label{sect:intro}

The study of the  characteristics of the strongly interacting dense matter inside neutron stars(NSs)\cite{Antoniadis:2013pzd,Fonseca:2021wxt,Riley:2021pdl,Miller:2021qha,Riley:2019yda,LIGOScientific:2017vwq,Radice:2017lry} is a topic of contemporary interest\cite{Baym:2017whm,Oertel:2016bki,Lovato:2022vgq}. It is believed that quark matter can exist inside the NS\cite{Annala:2019puf}. The first-principles methods however cannot be used  for describing quark matter at densities relevant inside stellar cores because of the sign problem in lattice Monte Carlo simulations at nonzero chemical potentials\cite{deForcrand:2009zkb} and that perturbative QCD being  only effective\cite{Kurkela:2009gj} at significantly higher densities. There have been various efforts to incorporate nonperturbative effects in increasingly sophisticated models since perturbative QCD is insufficient for addressing the quark matter EoS. Various phenomenological models have been used to study quark matter
recently, e.g., MIT Bag model\cite{chodes1974}, quark mass density-dependent model \cite{peng2001a,wen2005a,Chu_2014,Benvenuto95} the Richard potential model\cite{sinha2013strange}, NJL model\cite{nambu1961njl,Klevansky92,Hatsuda:1994pi,Buballa:2003qv,Buballa:1998pr,Lenzi:2010mz,Hanauske:2001nc,Wang:2020wzs,Pfaff:2021kse}, the perturbation model\cite{Fraga2001} , the field correlator method\cite{Plumari2013}, the quark-cluster model\cite{Xu2003}, and many other models. These models, to
some extent, have their origin in the  free-particle system. Strange quark matter(SQM) plays an important role in many interesting fields for example hot and dense matter in heavy ion collision, the structure of compact stars etc.  Ever since W. H. Witten suggested \cite{Farhi:1984qu,torres2013quark,Ferrer:2015}that the SQM would be absolutely stable even at absolute zero temperature, there has been a lot of interest in studying it.

We consider the simple MIT Bag model \cite{chodes1974,glendenning2012compact} for studying strange quark matter  which is 
a bulk matter phase with u, d, and s quarks in chemical equilibrium along with a minor fraction of electrons. The Bag constant (B) is added to the thermodynamical potential  of the free fermion system in order to reflect the quark confinement. The Bag pressure $B$ is actually the energy density difference between the perturbative vacuum and the true vacuum \cite{Burgio:2001mk,Burgio2002} and is often considered to  be constant in literature. Quark matter does not become asymptotically free immediately during or after
the phase transition in contrary to the 
MIT  Bag model's a priori assumption that quarks are free inside the bag. In order to overcome this problem, Raha et al\cite{CHAKRABARTY1989112} al introduced the medium effects via quark mass with a density-dependent quark mass model. A similar effect can be incorporated in the  Bag model through a density-dependent Bag pressure. It is well-known that the quarks at high densities, relevant to neutron stars or hybrid star cores, prefer asymptotic freedom \cite{Burgio:2001mk,Burgio2002}. This fact justifies that the Bag pressure be density dependent rather than being a constant. There are several studies on hybrid stars with density-dependent Bag models \cite{Sen:2021cgl,Sen:2022lig,suman2023a}. Therefore in the present 
work, we consider the  similar medium dependence of the Bag pressure via density. For the strange quark stars(SQS), the Bodmer-Witten conjecture states that the absolute stability of SQM  is determined in terms of matter-energy per baryon $\varepsilon/\rho_B$ at zero pressure being smaller than that of $^{56}Fe$ nucleus \cite{Farhi:1984qu,torres2013quark,Ferrer:2015}. 
From the  rules of thermodynamics, pressure is given by
\begin{equation}\label{eq:stabilty}
    P=\rho^2\frac{\partial}{\partial \rho}(\frac{\varepsilon}{\rho})
\end{equation}
According to the relation Eq.(\ref{eq:stabilty}) minimum of $\frac{\varepsilon}{\rho}$ should occur at zero pressure.
A recent study \cite{Lugones} of SQM using a density-dependent quark mass model solves the inconsistency problem of this model by using Canonical ensemble formalism.
With this formalism $(\frac{\varepsilon}{\rho})_{min}$ occur exactly at zero pressure point and also the equation of state is generated via Euler relation.
\begin{equation}\label{eq:euler_rel}
    \varepsilon=-P+\sum \mu_i\rho_i
\end{equation}
Generally, for the calculation of the equation of state  for SQM, Grand Canonical ensemble formalism is used,  If the density-dependent Bag pressure is used, then it is thermodynamically inconsistent as it violates the Euler relation Eq.~\eqref{eq:euler_rel}as described in detail in the formalism section~\ref{sec:forma}. Therefore we propose in this work to use the more appropriate
 chemical potential-dependent Bag pressure instead of a density-dependent Bag pressure in the Grand Canonical ensemble. If we want to use a density-dependent Bag pressure we have to use a Canonical ensemble where the Euler relation is valid. Our main focus in this work is to study the quark matter EoS in different ensembles with different forms of Bag pressure and thereby establish the thermodynamic consistency. Once the EoS of strange quark matter is ready, we have calculated the mass-radius diagram of the strange star using the same.

This paper is organized as follows. In Sec.~\ref{sec:forma} we give the detailed formalism of the equation of state in both ensembles using both density-dependent and chemical potential-dependent Bag pressures. In Sec.~\ref{sec:results} we show the numerical results. Finally, we summarise  in Sec.~\ref{sec:conclusion}.

\section{Formalism}\label{sec:forma}

In Ref.~\cite{Lugones},  the phenomenological density-dependent quark mass model has been revisited and the thermodynamical inconsistency  has been resolved within the Canonical ensemble formulation. This necessitates revisiting the MIT Bag Model as well where the same inconsistency is expected when one incorporates the density-dependent Bag pressure in Grand Canonical formalism. The main motivation of this work is to remove this inconsistency and reformulate the MIT Bag model in a thermodynamically consistent manner.  The medium effects can be incorporated through the Bag pressure by introducing  its dependence on  proper intensive parameter depending on the chosen ensemble. In Grand Canonical ensemble, the proper quantity is the chemical potential whereas in the Canonical ensemble density is the appropriate intensive quantity.  
\subsection{Grand Canonical }
\label{sec:Grandcanonical}
In the Grand Canonical Ensemble (GE), the thermodynamic potential depends on the chemical potential($\mu$), volume(V), and temperature(T) and all the thermodynamical quantities can be derived from the Grand Canonical potential $\Omega(\mu,V,T)$). We are considering cold neutron star which implies T=0.
\begin{itemize}
    \item \underline{Density Dependent Bag pressure :}
\end{itemize}  
In Grand Canonical ensemble, pressure is defined as 
\begin{equation}\label{eq:GC1}
    P=-(\frac{\partial \Omega}{\partial V})_{T,\mu}
\end{equation} 
where $\Omega$ is the grand-potential, 
 $\Omega_{GC}=\frac{{\Omega}}{V}$ is grand-potential per unit volume and $N=\rho V$. We can write Eq.(\ref{eq:GC1}) as
\begin{equation}\label{eq:GC2}
    P=-\left[\frac{\partial(\Omega_{GC} V)}{\partial(\frac{N}{\rho})}\right]_{T,\mu}=-\Omega_{GC}+\rho\left[\frac{\partial \Omega_{GC}}{\partial \rho}\right]_{T,\mu}
\end{equation} 
The density derivative term arises from the baryonic density dependence of the grand potential (as we are taking the Bag pressure to be density dependent). 
The thermodynamical potential for the quark matter in Grand Canonical ensemble  can be written as 
\begin{equation} \label{eq_GC3} 
\begin{aligned}
 \Omega_{GC}=&-\frac{1}{\pi^2}\sum_{f=u,d,s}\int_0^{k_{f}}\frac{k^4}{\sqrt{k^2+m_f^2}}+B(\rho)=\sum_{f=u,d,s}\Omega_f+B(\rho)
\end{aligned}
\end{equation} 
where $\Omega_f$ is written as \cite{glendenning2012compact} 
\begin{equation} \label{eq_H2} 
\begin{aligned}
 \Omega_f=&-\frac{1}{4\pi^2}[\mu_f\sqrt{\mu_f^2-m_f^2}(\mu_f^2-\frac{5}{2}m_f^2)\\&+\frac{3}{2}m_f^4 ln\left[\frac{\mu_f+\sqrt{\mu_f^2-m_f^2}}{m_f}\right]]
\end{aligned}
\end{equation} 
 where $\Omega_f$ represents the free quark matter grand potential. The interactions and medium effects are taken care of by the Bag pressure $B(\rho)$ \cite{Burgio:2001mk,Burgio2002}.
Hence pressure  in the Grand Canonical ensemble with density dependence Bag pressure  is as follows:  
\begin{equation}
    P_{GC}=-\Omega_{GC}+\rho\frac{\partial B(\rho)}{\partial\rho}
\end{equation} 
We consider  two forms  of the density dependence in the Bag pressure. 
\begin{itemize}
    \item \underline{Gaussian Form \cite{Burgio:2001mk,Burgio2002} :}
\end{itemize}
\begin{equation}
    B(\rho)=B_{as}+(B_0-B_{as})e^{\left[-\beta_{\rho}(\frac{\rho}{\rho_0})^2\right]}
\end{equation}  
Here $B_0$ is the value of B at $\rho=0$, $B_{as}$ is the value of B for asymptotic values of   $\rho$, $\beta_{\rho}$ is a control parameter and $\rho_0$ is the saturation density.
\begin{itemize}
    \item \underline{Hyperbolic Form \cite{suman2023a}:}
\end{itemize}
\begin{equation}
    B(\rho)=B_{as}+\frac{B_0}{2}(1-tanh(\frac{\rho-\bar{\rho}}{\Gamma_{\rho}})))
\end{equation} 
 $B_{as}$, $B_0$, $\bar{\rho}$ and $\Gamma_{\rho}$ are the free parameters in Hyperbolic form similarly.

The minimum of $\frac{\varepsilon}{\rho}$ occurs at zero pressure if we use the inverse Legendre transformation.
\begin{equation}\label{eq:ilegendre}
    \varepsilon_{GC}=\Omega_{GC}+\sum_i \mu_i \rho_i
\end{equation}

\begin{figure*}[htp]
    \centering
    \includegraphics[width=0.45\textwidth]{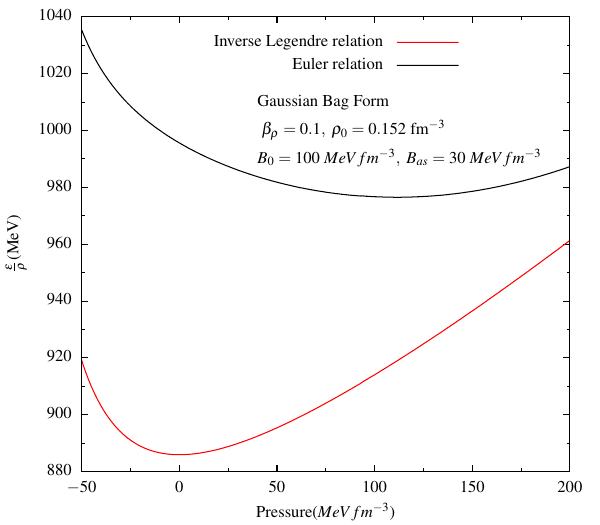}
    \includegraphics[width=0.45\textwidth]{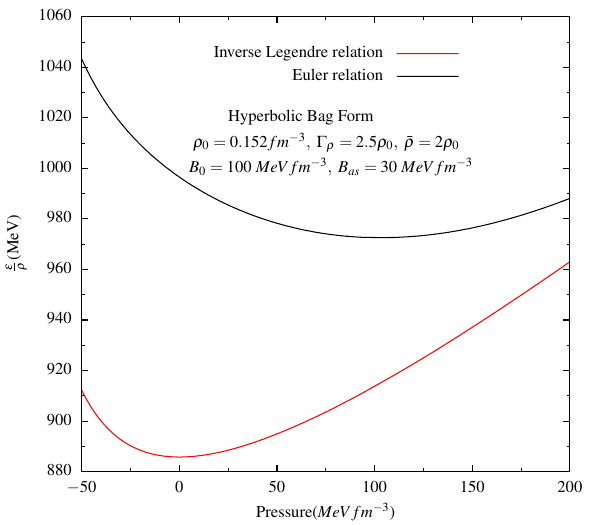}
  
   \caption{ Variation of energy per baryon with pressure using Euler relation and Inverse Legendre relation for Gaussian density-dependent(left) and  for Hyperbolic density dependent Bag (right) pressure.}
     \label{fig:euler}
\end{figure*} 

where $\epsilon$ is the energy density, $\mu_i$ is the chemical potential for the ith particle and $\rho_i$ is the density for the ith particle but the Euler relation (Eq.~\eqref{eq:euler}) is violated in this method. 
$\Omega_{GC}$ should be a function of $\mu$ but if  Bag pressure is taken to be density-dependent, then that is also not respected. So to avoid this kind of problem, we propose a proper thermodynamic treatment, where the medium effect is introduced via  a 
chemical potential dependent Bag pressure. 
\begin{itemize}
    \item \underline{Chemical potential dependent Bag pressure :}
\end{itemize} 
We consider similar depndence of chemical potential in the Bag pressure via Gaussian and Hyperbolic as in case of density dependent B.
The expression of B in the Grand Canonical are as follows : 
\begin{itemize}
    \item \underline{Gaussian Form :}
\end{itemize} 
\begin{equation}
    B(\mu)=B_{as}+(B_0-B_{as})e^{\left[-\beta_{\mu}(\frac{\mu}{\mu_0})^2\right]}
\end{equation}

\begin{itemize}
    \item \underline{Hyperbolic Form :}
\end{itemize}
\begin{equation}
   B(\mu)=B_{as}+\frac{B_0}{2}\left[1-tanh\left(\frac{\mu-\bar{\mu}}{\Gamma_{\mu}}\right)\right]
\end{equation} 
Here we are using a chemical potential-dependent Bag pressure, so the derivative term in Eq.~\eqref{eq:GC2} will not arise. Hence the pressure is
\begin{equation}\label{eq:pre_gc}
    P_{GC}=-\Omega_{GC}
\end{equation} 
The quark number densities are given by 
\begin{equation}\label{eq:density_modi}
    \rho_f=-\frac{\partial \Omega_{GC}}{\partial \mu_f}=\frac{k_f^3}{\pi^2}-\frac{\partial B(\mu)}{\partial \mu_f}
\end{equation} 
The density of each flavor is modified due to the chemical potential-dependent Bag pressure. This proper thermodynalical treatment where the thermodynamical potential in Grand Canonical ensemble depends on chemical potential and not density ensures that  the Euler relation is respected unlike the previous case.

\begin{equation}\label{eq:euler}
    \varepsilon_{GC}=-P_{GC}+\sum_f \mu_f \rho_f 
\end{equation} 
In the Grand Canonical ensemble, the energy from Eq.~\eqref{eq:euler}reads as
\begin{equation} \label{eq:gcen} 
\begin{aligned}
 \varepsilon_{GC}=\sum_f(\frac{1}{4\pi^2}&[(\mu_f^2-m_f^2)^{\frac{3}{2}}\mu_f-\frac{3}{2}m_f^2\mu_f\sqrt{\mu_f^2-m_f^2}\\&
 +\frac{3}{2}m_f^4 log (\frac{\mu_f+\sqrt{\mu_f^2-m_f^2}}{m_f})])+B(\mu)\\&
 +\sum_f\left[\mu_f(\frac{k_f^3}{\pi^2}-\frac{\partial B(\mu)}{\partial \mu_f})\right]
\end{aligned}
\end{equation}

\subsection{Canonical}
\label{sec:Canonical}

In the Canonical ensemble, all the thermodynamical quantities can be derived from the Helmholtz free energy $F(\rho,V,T)$. At zero temperature(T=0) 
\begin{equation}\label{eq:ce1}
    F(\rho,V)=U=\varepsilon V
\end{equation} 
where $U$ is the  total internal energy and $\varepsilon$ is the internal energy per unit volume.
 The Helmholtz free energy per unit volume is given by 
\begin{equation}\label{eq:free_en} 
    \varepsilon_C=\sum_{f=u,d,s}\frac{3}{\pi^2}\int_0^{k_f} k^2\sqrt{k^2+m_f^2}dk+B(\rho) 
\end{equation} 
where the medium effects is incorporated in Bag pressure by making it density dependent since density is the appropriate intensive parameter in Canonical ensemble.
We can rewrite the Eq.~\eqref{eq:free_en} as 
\begin{equation} \label{eq:ce3}
\begin{aligned}
 \varepsilon_C=\sum_f(\frac{3}{4\pi^2}&[k_f^3\sqrt{k_f^2+m_f^2}+\frac{1}{2}m_f^2k_f\sqrt{k_f^2+m_f^2}\\&
 -\frac{1}{2}m_f^4 log \left(\frac{\sqrt{k_f^2+m_f^2}+k_f}{m_f}\right)])+B(\rho)\\&
\end{aligned}
\end{equation} 
Unlike the case of  energy in Grand Canonical ensemble Eq.~\eqref{eq:gcen}, the expression for energy in Canonical  does not have any derivative dependent term of Bag pressure.   As in the case of Grand Canonical ensemble, here also two forms for the density dependence is being considered. 
Particle number density is given by 
\begin{equation}\label{eq:ce5} 
    \rho_f=\frac{ k_f^3}{\pi^2}
\end{equation} 
Total baryon no density is
\begin{equation}\label{eq:ce6} 
    \rho=\frac{1}{3}(\rho_u+\rho_d+\rho_s)
\end{equation} 
The energy density at T=0 is $\epsilon_c=\frac{F}{V}$
\begin{equation}\label{eq:ce7} 
    \varepsilon_C=\sum_f \epsilon_f+B(\rho)=\sum_f\frac{F_f}{V}+\frac{F_B}{V}
\end{equation}
where $F_f$ is the free energy for the contribution for the free fermions and $F_B$  is the free energy contribution from the Bag pressure.
\begin{equation}\label{eq:ce8} 
    P_C=-\frac{\partial F}{\partial V}=-\frac{\partial }{\partial V}\left(\sum_f F_f+F_B\right)
\end{equation} 
Pressure in this ensemble is given by
\begin{equation}\label{eq:ce13} 
       P_C=\left(\sum_f \rho_f^2\frac{\partial }{\partial \rho_f}(\frac{\epsilon_f }{\rho_f})\right)+\rho^2\frac{\partial }{\partial\rho}\left(\frac{ B(\rho)}{\rho}\right) 
\end{equation}
\begin{equation}\label{eq:ce14} 
    P_C=\sum_f P_f+P_{Bag}
\end{equation}
where, 
\begin{equation} \label{eq:ce16} 
\begin{aligned}
 P_f=\frac{1}{4\pi^2}(k_f^3\sqrt{k_f^2+m_f^2}+\frac{1}{2}m_f^2k_f\sqrt{k_f^2+m_f^2}&\\-\frac{1}{2}m_f^4 log \left(\frac{\sqrt{k_f^2+m_f^2}+k_f}{m_f}\right))\\
\end{aligned}
\end{equation} 
\begin{equation}
    P_{Bag}=-B(\rho)+\rho\frac{\partial B(\rho)}{\partial \rho}
\end{equation}
Therefore pressure can be rewritten as 
\begin{equation}\label{eq:ce17} 
    P_C=\frac{1}{\pi^2}\sum_{f=u,d,s}\int_0^{k_f}\frac{k^4}{\sqrt{k^2+m_f^2}}-B(\rho)+\rho\frac{\partial B(\rho)}{\partial\rho}
\end{equation} 
Unlike the case of Grand Canonical Eq.~\eqref{eq:pre_gc}, pressure here has a derivative term of $B(\rho)$. 
The chemical potential is
\begin{equation}\label{eq:cano_mu} 
    \mu_f=\frac{\partial F}{\partial N_f}=\frac{\partial\varepsilon_f}{\partial \rho_f}+\frac{\partial B(\rho)}{\partial \rho_f}=\sqrt{k_f^2+m_f^2}+\frac{\partial B(\rho)}{\partial \rho_f}
\end{equation} 
In this ensemble quark chemical potential is modified due to the medium effects in MIT 
 Bag model through the  density dependence.  A similar effect is observed in a Grand Canonical ensemble where the density is modified due to the chemical potential dependent Bag pressure. This modified chemical potential satisfies the Euler relation thereby establishing the validity of the method. 
\section{Results}\label{sec:results}
We focus on quark matter that might exist inside a NS. For the SQM, we include  electrons also and we have to take into account  the chemical equilibrium condition, $\mu_d=\mu_u+\mu_e=\mu_s$, the charge neutrality condition, $\frac{2}{3}\rho_u-\frac{1}{3}\rho_d-\frac{1}{3}\rho_s-\rho_e=0$ and baryon no density conservation $\rho=\frac{1}{3}(\rho_u+\rho_d+\rho_s)$. In our calculation, we take $m_u=5.0$~MeV, $m_d=7.0$~MeV and $m_s=95.0$~MeV
\cite{ParticleDataGroup:2020}.
\subsection{ Grand Canonical ensemble} 
\begin{figure*}[htp]
    \centering
    \includegraphics[width=0.45\textwidth]{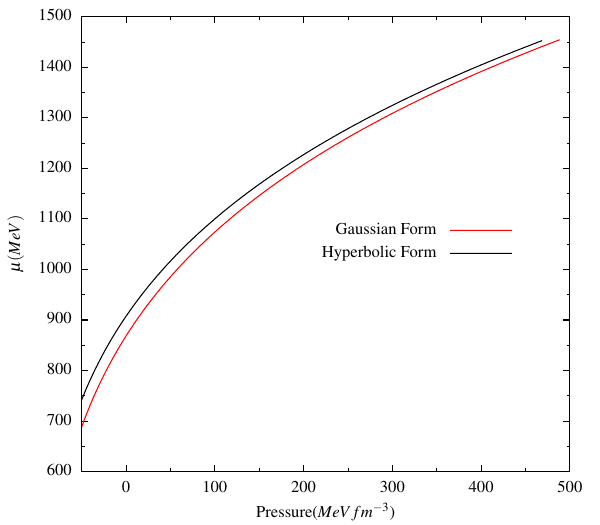}
    \includegraphics[width=0.45\textwidth]{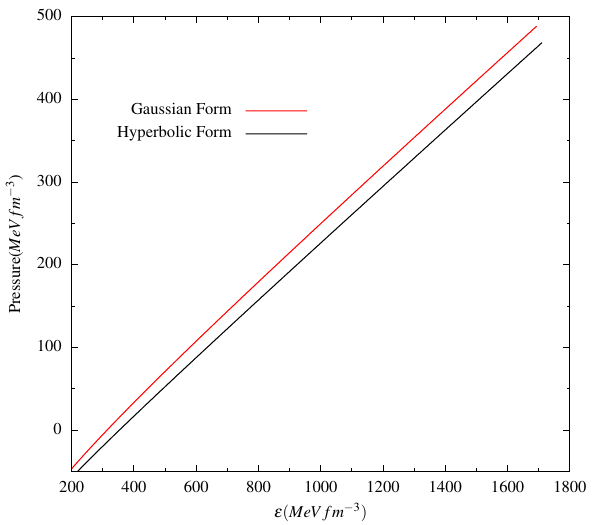}
    \includegraphics[width=0.45\textwidth]{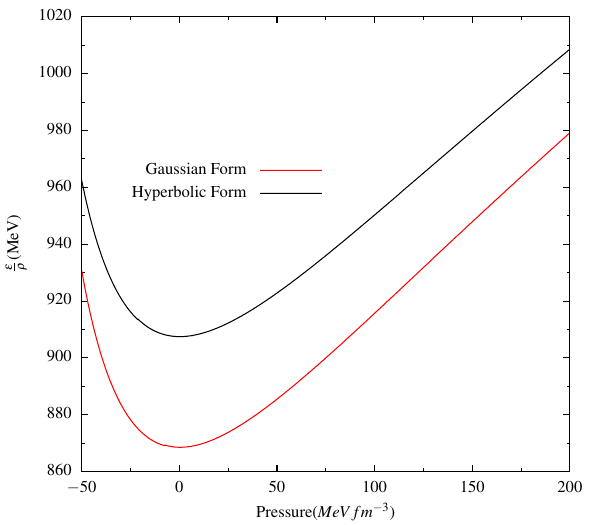}
    \includegraphics[width=0.45\textwidth]{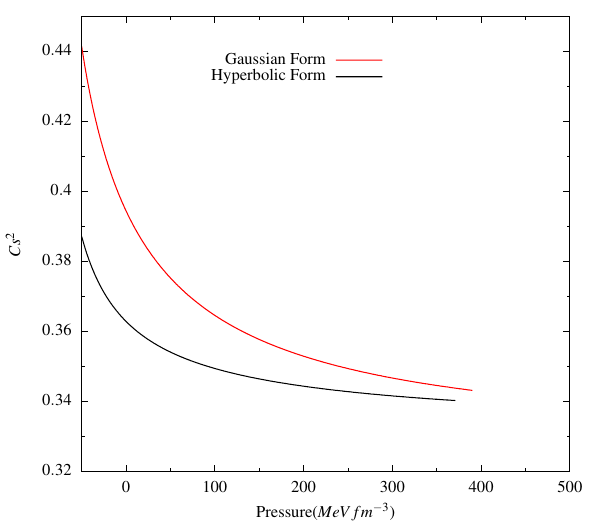}
  \caption{Equation of state and speed of sound in the Grand Canonical ensemble for Gaussian and Hyperbolic chemical potential dependent Bag pressure, variation of chemical potential with pressure, variation of pressure with energy density, variation of  $\frac{\varepsilon}{\rho}$ with pressure and variation of   speed of sound with pressure {{with $B_0$=100 MeV $fm^{-3}$},~$B_{as}=30$ MeV $fm^{-3}$,~$\beta_{\mu}=1.0$,~$\mu_0=1000 MeV$,~$\Gamma_{\mu}=1000 MeV$,~$\bar{\mu}=800 MeV$} } 
\label{fig:gc}
\end{figure*}  

\begin{figure*}[htp]
    \centering
    \includegraphics[width=0.45\textwidth]{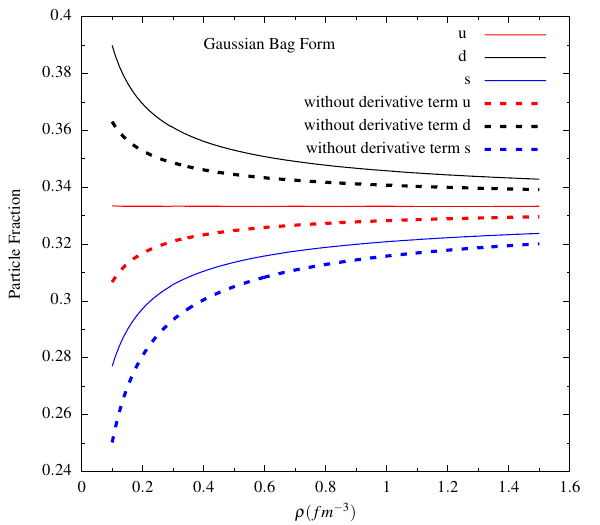}
     \includegraphics[width=0.45\textwidth]{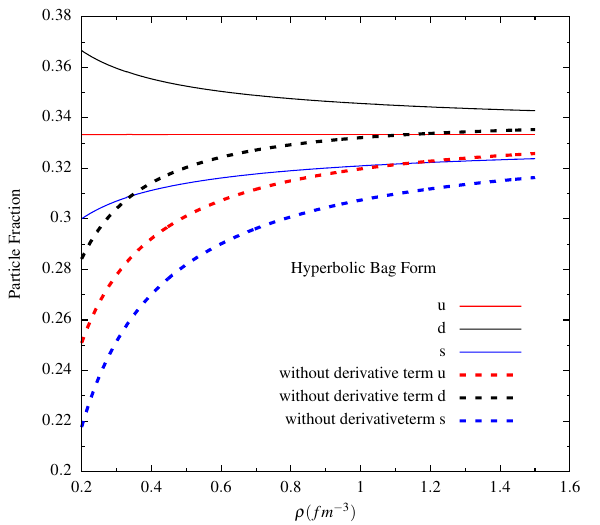}
    \caption{Particle fraction in the Grand Canonical ensemble for Gaussian (left) and Hyperbolic (right) chemical potential dependent Bag pressure.}
     \label{fig:part_frac}
     
\end{figure*}

First, we study the thermodynamic stability condition for the SQM in a Grand Canonical ensemble with density-dependent Bag pressure shown in Fig.~\ref{fig:euler}. From Eq.~\eqref{eq:stabilty}, 
the minimum of $\frac{\varepsilon}{\rho}$ should occur at zero pressure. In Fig.~\ref{fig:euler} we use both Gaussian and Hyperbolic density-dependent Bag pressure. In Fig.~\ref{fig:euler} we see that if we use Euler relation Eq.~\eqref{eq:euler} then minimum of $\frac{\varepsilon}{\rho}$ does not occur at zero pressure but if  inverse Legendre transformation Eq.~\eqref{eq:ilegendre} is used,  then minimum occurs at zero pressure but it violates the  rules of thermodynamics.

In order to establish thermodynamic consistency,  we propose a chemical potential-dependent Bag pressure, where  self-consistent thermodynamic treatment is being restored. We study the equation of state in a Grand Canonical ensemble with both Gaussian and Hyperbolic chemical potential dependent Bag pressure.  For this, we take $B_0=$100 MeV$fm^{-3}$ and $B_{as}=$30 MeV$fm^{-3}$, same for both Gaussian and Hyperbolic form. For Gaussian $\beta_{\mu}=1.0$,$\mu_0=1000$MeV and for Hyperbolic $\bar{\mu}=800MeV$ and $\Gamma_{\mu}=1000MeV$. In Fig.~\ref{fig:gc} upper panel, variation of chemical potential with pressure and variation of pressure with energy density is being displayed. We see that the chemical potential has a similar kind of variation with energy density both for the Gaussian and Hyperbolic case; the equation of state ( pressure versus energy density) also has a similar pattern for both forms of the Bag pressure. In the lower-left panel, we show that the minimum of $\frac{\varepsilon}{\rho}$ exactly occurs at zero pressure which will not happen in the case of density-dependent Bag pressure obeying the Euler relation. 
 In the lower right panel, variation of the speed of sound with pressure is shown; due to chemical potential dependent Bag pressure, $C_s^2$  varies with pressure and it tends to  $\frac{1}{3}$ for higher  values of pressure.
 
 In Fig.~\ref{fig:part_frac}, we show  the variation of  particle fraction variation with density; the solid line represents the particle fraction according to Eq.~\eqref{eq:density_modi}
  and the dotted line represents the particle fraction without taking the derivative of the Bag pressure as in right hand side of Eq.~\eqref{eq:density_modi}. 
  Particle fraction is modified due to the derivative term in the Eq.~\eqref{eq:density_modi}, and since Bag pressure varies with  chemical potential,  it reflects in the particle fraction plot.

 Bodmer-Witten conjecture\cite{Farhi:1984qu,torres2013quark,Ferrer:2015} states that for  SQM to be  stable and be the true ground state of matter the stability  is dictated by energy density per baryon, which is regulated by Bag pressure.   The allowed range of the different parameters ($B_0, B_{as},\beta_{\mu}$) in the $\mu$ dependent Bag pressure(Gaussian form) have been estimated as per Bodmer-Witten conjecture  as shown in Table~\ref{tab:grandcano_stab}. The upper limit of $\beta_{\mu}$ is estimated for 2-flavour quark matter, when $\frac{\varepsilon}{\rho}\ge 930$ and lower limit of $\beta_{\mu}$ is estimated for 3-flavour quark matter, when $\frac{\varepsilon}{\rho}\le 930$. In a similar way, we can estimate the stability range  of the parameters in the Hyperbolic Bag pressure which is not shown here for the sake of brevity.
\begin{table} [h!]
\centering
\caption{$\mu$~dependent Gaussian Bag pressure in Grand-Canonical ensemble}
\begin{tabular}{|c|c|c|}
\hline \hline
 $B_0$ (MeV $fm^{-3}$) & $B_{as}(MeV fm^{-3})$ & $\beta_{\mu}$ \\
\hline
 60& 25 &[0.0,0.12] \\
 \hline
 100& 50 &[0.48,2.33] \\
 100& 40 &[0.37,1.45] \\
 100& 30 &[0.31,1.11] \\
 \hline
  150& 50 &[1.30,3.10] \\
  150& 40 &[1.08,2.15] \\
  150& 30 &[0.93,1.71] \\
   \hline
  200& 50 &[1.75,3.60] \\
  200& 40 &[1.50,2.55] \\
  200& 30 &[1.35,2.10] \\
\hline
\end{tabular}
\label{tab:grandcano_stab}
\end{table}  
 
\subsection{Canonical ensemble}\label{subsec:results_cano}

\begin{figure*}[htp]
    \centering
    \includegraphics[width=0.45\textwidth]{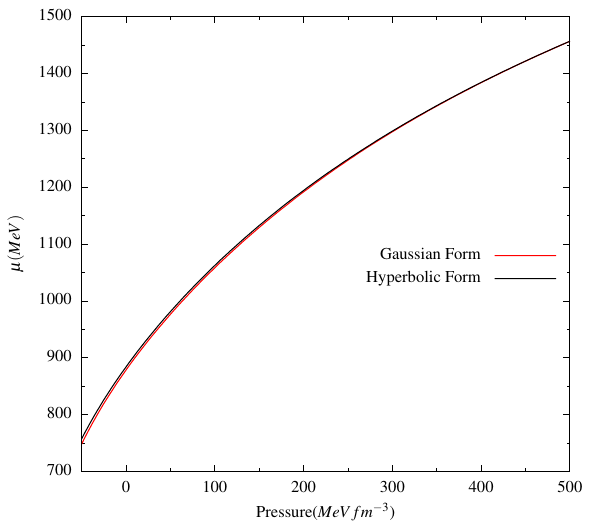}
    \includegraphics[width=0.45\textwidth]{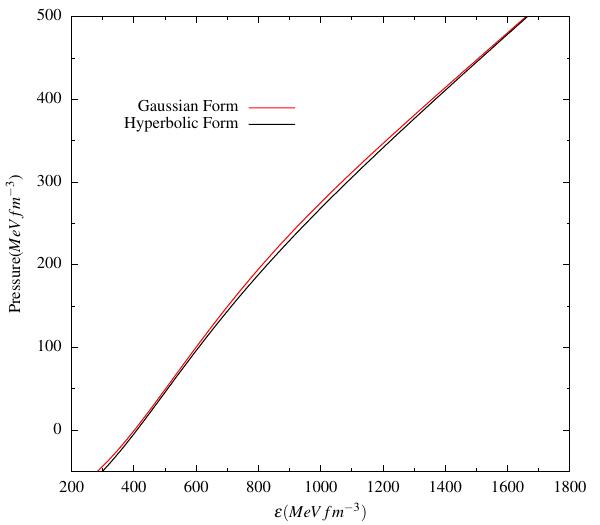}
    \includegraphics[width=0.45\textwidth]{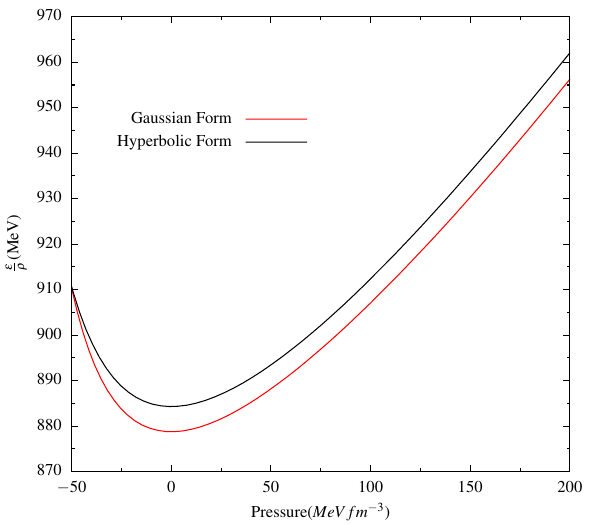}
    \includegraphics[width=0.45\textwidth]{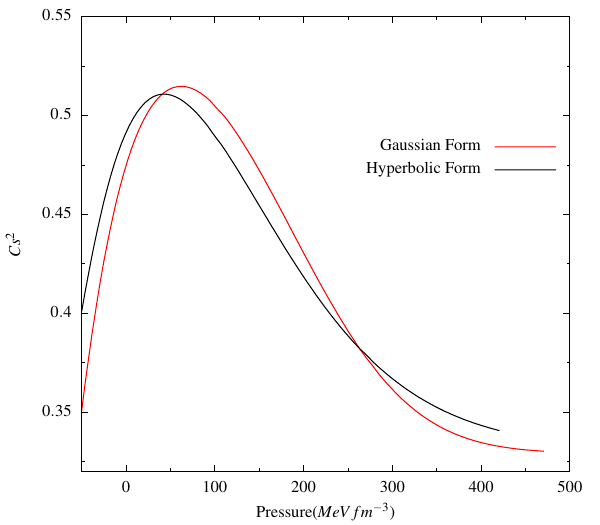}
   \caption{Equation of state and speed of sound in the Canonical ensemble for Gaussian and Hyperbolic density dependent Bag pressure, variation of chemical potential with pressure, variation of pressure with energy density, variation $\frac{\varepsilon}{\rho}$ with pressure and variation speed of sound with pressure with $B_0$=100 MeV $fm^{-3}$,~$B_{as}=30$ MeV $fm^{-3}$,~$\beta_{\rho}=0.1$,~$\rho_0=0.152 fm^{-3}$,~$\Gamma_{\rho}=2.5\rho_0$,~$\bar{\rho}=2\rho_0$} 
     \label{fig:cc}
\end{figure*}

In the Canonical ensemble, the medium effects are taken into account through a density-dependent Bag pressure. This ensures thermodynamic consistency and validity of Euler's relation.  For this, we take $B_0=$100 MeV$fm^{-3}$ and $B_{as}=$30 MeV$fm^{-3}$, same for both Gaussian and Hyperbolic form. For Gaussian $\beta_{\rho}=0.1$,$\rho_0=0.152~fm^{-3}$ and for Hyperbolic $\bar{\rho}=2\rho_0$ and $\Gamma_{\rho}=2.5\rho_0$. In Fig.~\ref{fig:cc} upper panel shows the variation of chemical potential with pressure and the variation of pressure with energy density. It is observed that these two variations are similar and close for the Gaussian and the Hyperbolic  types variation of the density-dependent Bag pressure. In the lower-left panel, we show that the minimum of $\frac{\varepsilon}{\rho}$ exactly occurs at zero pressure which will not happen in the case of density-dependent Bag pressure in Grand Canonical ensemble( Fig.~\ref{fig:euler})obeying the Euler relation). Therefore if we want to use a density-dependent medium effect, we have to use a Canonical approach.  
 In the lower right panel, variation of the speed of sound with pressure  due to the density-dependent medium effect is displayed. It is observed that $C_s^2$  tends to conformal limit at higher pressure. The variation of $C_s^2$ at lower values of pressure differs from that of the Grand Canonical since energy(See Eq.~\eqref{eq:gcen} and Eq.~\eqref{eq:ce3}) and pressure(See Eq.~\eqref{eq:pre_gc} and Eq.\eqref{eq:ce17}) differs in both the ensembles.

\begin{figure*}[htp]
    \centering
    \includegraphics[width=0.45\textwidth]{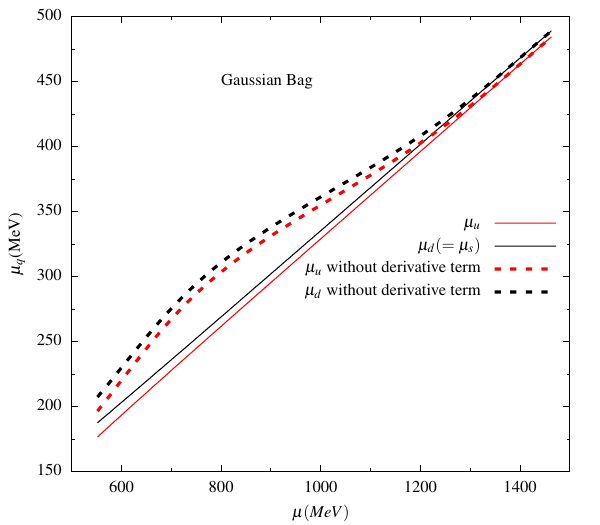}
     \includegraphics[width=0.45\textwidth]{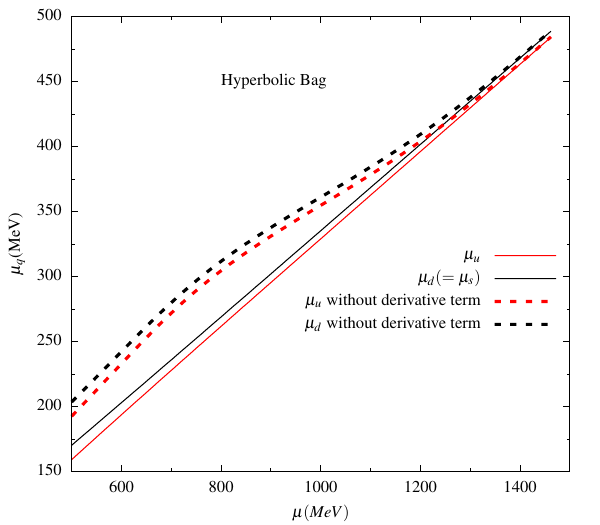}
  \caption{Quark chemical potential in the Canonical ensemble for Gaussian (left) and  Hyperbolic (right)density dependent Bag pressure.}
\label{fig:chemical_pot}  
\end{figure*}

  In Fig.~\ref{fig:chemical_pot},  the quark chemical potential variation with baryonic chemical potential is being shown;  here the solid line represents the individual chemical potential of different quarks (u,d,s), and the dotted line represents the  same chemical potential without taking the derivative of the Bag pressure term as shown in Eq.~\eqref{eq:cano_mu}. At higher values of the chemical potential, the Bag pressure is almost constant, therefore the derivative term Eq.~\eqref{eq:cano_mu} does not contribute and hence the lines merge as seen in Fig.~\ref{fig:chemical_pot}. 

 Following the same procedure as previously stated in case of the Grand Canonical ensemble, the allowed free parameters in the $\rho$ dependent Bag pressure( Gaussian form)  constrained by Bodmer-Witten conjecture ($B_0, B_{as},\beta_{\rho}$) is estimated as shown in Table~\ref{tab:cano_stab}.
 In a similar way, we can estimate the stability limit of the parameters following the Hyperbolic Bag pressure.
   
\begin{table} [h!]
\centering
\caption{$\rho$~dependent Gaussian Bag pressure in Canonical ensemble}
\begin{tabular}{|c|c|c|}
\hline \hline
 $B_0$ (MeV $fm^{-3}$) & $B_{as}(MeV fm^{-3})$ & $\beta_{\rho}$ \\
\hline
57& 40 &[0.0,0.013] \\
\hline
 100& 50 &[0.052,0.59] \\
 100& 40 &[0.041,0.33] \\
 100& 30 &[0.035,0.23] \\
 \hline
  150& 50 &[0.113,0.75] \\
  150& 40 &[0.091,0.44] \\
  150& 30 &[0.077,0.31] \\
   \hline
  200& 50 &[0.141,0.85] \\
  200& 40 &[0.115,0.50] \\
  200& 30 &[0.098,0.36] \\
\hline
\end{tabular}
\label{tab:cano_stab}
\end{table} 

\subsection{Mass-Radius diagram}
\begin{figure*}[htp]
    \centering
\includegraphics[width=0.45\textwidth]{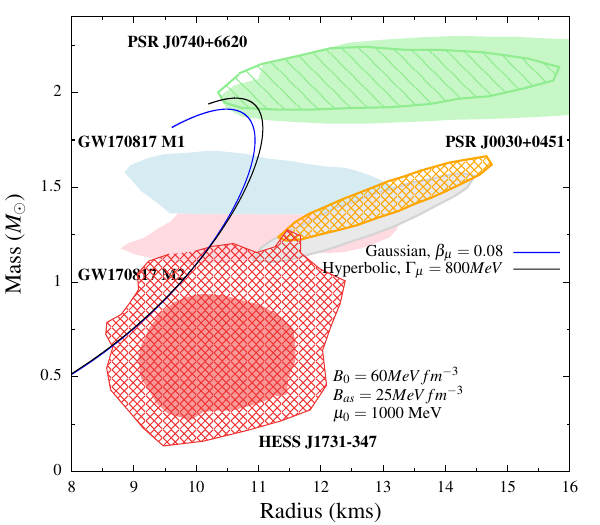}
     \includegraphics[width=0.45\textwidth]{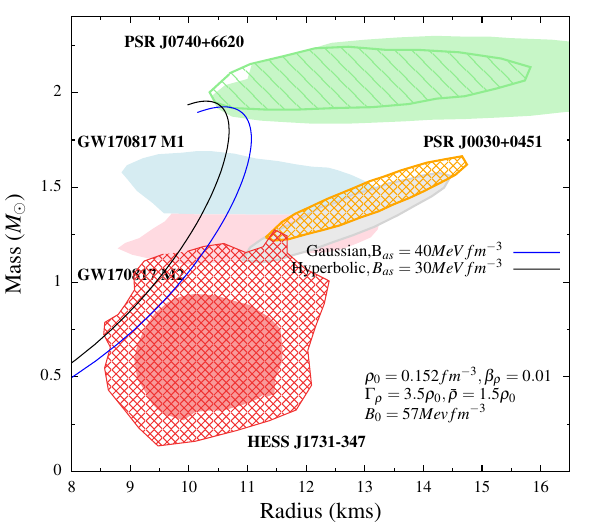}
    \caption{ Mass-radius relationship of quark star with chemical potential dependent values of bag pressure for Gaussian form of $B(\mu)$ and Hyperbolic form $B(\mu)$ (left) and density dependent values of bag pressure for Gaussian form of $B(\rho)$ and Hyperbolic form $B(\rho)$. Observational limits imposed from HESS J1731-347  \cite{2022NatAs} and the constraints on $M-R$ plane prescribed from GW170817 \cite{LIGOScientific:2018cki}) and   PSR J0740+6620\cite{Fonseca:2021wxt} are also compared.}

  \label{fig:mr}  
\end{figure*}

Exploring the inner structures of strange quark stars will be simple once the EoS of strange quark matter is calculated. Therefore we present mass-radius configuration in Fig.~\ref{fig:mr}. We use $\mu$ dependent Bag pressure  in the Grand Canonical ensemble in both Gaussian and Hyperbolic forms and the chosen parameters are given in the left side of Fig.~\ref{fig:mr}. 
  We have  also used $\rho$ dependent Bag pressure in the Canonical ensemble in both Gaussian and Hyperbolic forms and the chosen parameters are given on the right side of Fig.~\ref{fig:mr}. Different parameters of the MIT Bag model used in the mass-radius diagram are chosen such that they satisfy the Bodmer Witten conjecture \cite{Farhi:1984qu} for stability of stars  as well as the recent astrophysical data that are shown in Fig.~\ref{fig:mr}. The calculated quark star configurations satisfy the recently obtained constraint from the low-mass compact object HESS J1731-347 and constraints on the mass and radius of compact stars from  GW170817. The maximum mass constraint from PSR J0740+662 is satisfied for the $\mu$ dependent hyperbolic case only, though it is  dependent on the parameter space. Our main focus in this work is  the proper thermodynamic treatment and stability; in future, one can explore more on quark star and hybrid star properties with chemical potential dependent $B(\mu)$(varying the parameters) and compare those with  different  astrophysical constraints.
\section{SUMMARY AND CONCLUSION}\label{sec:conclusion}
In this work, we have studied the medium effects of quark matter through the Bag pressure in the framework of MIT Bag model. We demonstrate that if a density-dependent Bag pressure is used in the Grand Canonical ensemble, then in the equation of state,  either the Euler relation (Eq.~\eqref{eq:euler_rel}) becomes invalid or the  lowest energy per baryon does not coincide with that of zero pressure. In order  to overcome this inconsistency in thermodynamics, we suggest  that the medium influence of SQM to be incorporated via density or chemical potential dependent  Bag pressure depending on the ensemble chosen. In other words, the intensive parameter which addresses the medium effects should be ensemble dependent. This work is one of the first to propose chemical potential dependent Bag pressure in Grand Canonical ensemble .  The self-consistency in thermodynamics is restored if the chemical potential dependent Bag parameter is used in the Grand Canonical ensemble and density dependent in the Canonical ensemble. In the Grand Canonical ensemble, density is modified due to chemical potential-dependent Bag pressure whereas in the Canonical ensemble, chemical potential is modified due to the density-dependent Bag pressure. The main ingredient of our study is the reformulation of medium-dependent Bag pressure according to the  choice of ensemble which solves the  inconsistency problem from thermodynamics point of view. Generally, for the infinite matter system, Grand Canonical ensemble is used. If one chooses to take the density-dependent medium effect, then Canonical is the appropriate ensemble. We have calculated the mass radius (M-R) of strange stars using this thermodynamically  consistent formalism. This can be further explored for the study of the equation of state and different structural properties of the strange stars and  the hybrid stars in future.

\bibliography{mitbag}

\end{document}